\documentclass[conference,table]{IEEEtran} 
\IEEEoverridecommandlockouts

\usepackage[acronym]{glossaries}
\usepackage{hyperref}
\usepackage{multicol}
\usepackage{chronology}
\graphicspath{{figures/}}

\usepackage{adjustbox}

\usepackage{booktabs} % for the \toprule et al. to make tables look better.
\usepackage{array} % needed for tabular's 'm' 
\newcolumntype{L}[1]{>{\raggedright\let\newline\\\arraybackslash\hspace{0pt}}m{#1}} % https://tex.stackexchange.com/questions/12703/how-to-create-fixed-width-table-columns-with-text-raggedright-centered-raggedlef

\usepackage{tikz}
\usetikzlibrary{positioning}

\usepackage[autostyle]{csquotes}

\usepackage{soul}
\newcommand{\attrivute}[1]{
	{\hfill (\scriptsize #1)}\newline
}

\newcommand{\fc}[1]{
	\attrivute{#1}
	\cellcolor{gray!15}
}

\newcommand{\f}[1]{
	\attrivute{#1}
}

\newcommand{\m}[1]{
	\attrivute{#1}
	\cellcolor{red!15}
}

\newcommand{\acs}[1]{\acrshort{#1}}
\newcommand{\acl}[1]{\acrlong{#1}}
\newcommand{\acf}[1]{\acrfull{#1}}

\makeglossaries
\newacronym{attk}{ATT\&CK}{MITRE ATT\&CK™}
\newacronym{ics}{ICS}{Industrial Control System}
\newacronym{ioc}{IOC}{Indicator Of Compromise}
\newacronym{iocs}{IOCs}{Indicators Of Compromise}

\begin{document}

\title{Big Fish, Little Fish, Critical Infrastructure: An Analysis of Phineas Fisher and the `Hacktivist' Threat to Critical Infrastructure}

\author{
	\IEEEauthorblockN{Peter Maynard and Kieran McLaughlin\\
	Centre for Secure Information Technology\\
	Queen's University Belfast, UK\\
	\{p.maynard,kieran.mclaughlin\}@qub.ac.uk}
}

\maketitle

\begin{abstract}
The hacktivist threat actor is listed in many risk decision documents. Yet their tactics and techniques often remain a mystery. We create a \acf{attk} model of a well known hacktivist who goes under the pseudonym of Phineas Fisher, and map that threat to critical infrastructure. The analysis is derived from hacker manifestos, journalist reporting, and official government documentation. This analysis fills a gap in current threat models, to better define what skills and methods a determined hacker might employ. This paper also identifies seven essential mitigations which can be deployed by critical infrastructure operations and asset owners, to prevent such intrusions by hacktivists. We are in the process of contributing this threat actor into the \acl{attk} knowledge base. 
\end{abstract}


\section{Introduction}

State actors are widely considered to be the default threat actors to critical infrastructure, since they are the threat most often discussed in the media. For example, in 2017 the malware CrashOverride \cite{eset_industroyer_2017} was found to have been used in the 2016 power outages in Ukraine, which saw the manipulation of industrial control devices, resulting in a few hours of downtime in a localised area. A report published in 2019 \cite{slowik_crashoverride_2019} suggests that their objective was to disable the whole country's power grid for a much longer period. While it may make little practical difference to an operator who the intruder is, it is good practice to have a solid comprehension of what tactics and techniques an adversary may deploy. In this instance, the creators had time and money to devote to the development of a sophisticated piece of malware, specifically designed to compromise the target. The skills and techniques they deployed are well understood, and attacks like these are becoming more common. However, the techniques and tactics used by less resourced actors often go undiscussed.

State of the art literature \cite{ghafir_security_2018,white_risk_2019} does not detail the technical threats which a hacktivist might employ, since it is hard to attain such information without identifying and investigating the intruder. We, therefore, perform an analysis of a well known hacktivist who goes under the pseudonym of Phineas Fihser (See section \ref{sec:phineas}). The majority of this analysis is based on their self-published manifestos, which break down the steps taken to compromise their targets. This analysis is further supported by news reporting and academic literature. Subsequently, we derive a \acf{attk} model of their techniques, and identify ways that an equivalently capable hacktivist like Fisher might compromise critical infrastructure. Based on this research, we identify seven mitigations which may be deployed within an industrial network. This work was motivated by a recent publication by Fisher in November 2019, which stated `I will pay up to 100 thousand USD for each filtration of this type, according to the public interest and impact of the material...'\cite{fisher_hackback_2019}, where they advocate the intrusion into oil, gas, mining, logging and livestock companies, and surveillance companies such as the NSO group, among others. While we do not intend to make judgements on the political aims of such `hacktivism', based on this call to arms, and lack of insight into hacktivist methods, it is clear there is a need for further technical analysis. 

The remainder of the paper provides a discussion on the related work (\S\ref{sec:related}), followed by an introduction of the \acs{attk} model (\S\ref{sec:attk}) and the Hacktivist known as Phineas Fisher (or Phisher) (\S\ref{sec:phineas}). We then offer an analysis of Fisher's attacks (\S\ref{sub:anlysis-phisher}) and how this threat may be mapped to critical infrastructure (\S\ref{subsec:threat}), followed by proposed mitigations (\S\ref{subsec:mitigations}) and concluding remarks (\S\ref{sec:conclusion}).

\section{Related Work}
\label{sec:related}

% Threats to CNI
To date, several studies have investigated threats to critical infrastructure and industrial control systems. One such paper, by Rudner \cite{rudner_cyber-threats_2013}, identified several threat actors: international terrorism, state-sponsored terrorism, espionage and sabotage, malevolent hacktivism, and insider threats. Rudner examines the declared intentions, strategies, objectives and demonstrated capabilities of those entities known to have threatened Critical National Infrastructure. These threats align with the NIST SP-800-82 Guide to industrial control system security \cite{stouffer_guide_2015} definitions, as four primary adversary actors: Individual; Group; Organisation; and Nation-State. Yet, neither of these publications define what actions these actors may take against their targets.

The most interesting threat, regarding our investigation, is `malevolent hacktivism', which Rudner cites the United States Department of Homeland Security warning that Anonymous may target critical infrastructure \cite{department_of_homeland_security_dhs_2011} "as part of its green energy agenda", which specifically supports the environmentalist campaign against the Alberta Oil Sands and the proposed Keystone XL oil pipeline. Two other groups are listed (Deep Green Resistance and Fertile Ground), who in 2011/12 declared an intention to target critical infrastructure. Fertile Ground propounded\footnote{Based on private communication from a senior security officer in the Canadian energy industry, January 2012 \cite{rudner_cyber-threats_2013}.} the view that critical infrastructure is highly vulnerable and poorly designed, so that cyber attacks striking at key nodes could have a significant impact. As of early 2020, the authors have been unable to find any incidents that could be attributed to them. 

While a variety of incidents have been attributed to Anonymous, none appear to (or have been publicly reported to) directly affect critical infrastructure. Meanwhile, Anonymous has claimed \cite{yaron_hackers_2011} they have access to the Stuxnet source code, but there has been no evidence that they have used it. Moreover, Stuxnet was designed to run on a specific site and is not particularly useful on its own.

% Technical Skills of a hacktivist | Human Threats to CI 
The human threat to critical infrastructure is discussed by Ghafir \cite{ghafir_security_2018}, in which they propose a system to improve the security awareness of business environment employees. Of particular interest is Ghafir's discussion of social engineering and the attack strategies, suggesting the use of Kevin Mitnick's attack cycle, i.e. Research; Develop Trust; Exploit Trust, and; Utilise Information. In emergencies where many disperse departments and business partners all need to interoperate, social engineering becomes a very valid threat to critical infrastructure. Indeed, as reported by \cite{ics-cert_incident_2012}, spear-phishing is a common entry point. Nonetheless, the security awareness delivery method proposed by Ghafir does not appear particularly suited to application to SCADA/ICS systems, due to the addition of `pop-ups' to employees workflows along with additional network connections. 

%  Current work is missing 
Generally, the current body of research describes intrusions at a very abstract level, and primarily focus on motivation and description of the different types of threat actors. There is a lack of detailed technical analysis of the skills a hacktivist may employ when compromising critical infrastructure. One might contemplate there is no difference between a hacktivist or state actor, and since there is no publicly attributed attack to critical infrastructure that has been performed by an individual or group of hacktivists, one could assume they may follow the existing intrusion trends as reported by ICS-CERT. This paper aims to explore these assumptions. 

\section{\acl{attk}}
\label{sec:attk}

% MITRE ATTACK

Released in 2015, \acf{attk} \cite{strom_finding_2017} is a curated knowledgebase of adversary behaviours. \acs{attk} has three main corpora consisting of pre-ATT\&CK, mobile, and enterprise. This paper considers the enterprise version since it is designed for Microsoft Windows and Linux based operating systems. The knowledge base consists of adversary tactics (why) and techniques (how), that can be used by defenders to determine how secure their systems are. Tactics serve as useful contextual categories for individual techniques and cover standard notations for things adversaries do during an operation, such as persist, discover information, move laterally, execute files, and exfiltrate data. Techniques represent how an adversary achieves a tactical objective by performing an action.
% abstraction levels | grounded | other models 
As detailed in \cite{strom_mitre_2018}, \acs{attk} has multiple applications:  Adversary Emulation; Red Teaming; Behavioural Analytics Development; Defensive Gap Assessment; SOC Maturity Assessment; Cyber Threat intelligence Enrichment. This paper will use the model to analyse the threat of Fisher to critical infrastructure, in a sense a gap assessment will be performed based on the hacktivist threat.  

% Other models

\begin{figure}
	\centering
  	% arara: pdflatex 

% \documentclass{standalone}
% \usepackage{tikz}
% \usetikzlibrary{positioning}
% \begin{document}

\begin{tikzpicture}[
	every node/.style={
		rectangle,
		draw=black!60,
		thick,
		inner sep=5pt,
		text width=5.5cm,
		align=center},
	absLab/.style={
		draw=none,
		text width=.6cm,
	},
	absLab1/.style={
		draw=none,
		text width=3.2cm,
	},
	node distance=1.5mm,
]

\node (high) {High-level Models\\(Lockheed Martin KillChain$^{®}$,\\Microsoft STRIDE)};
\node (mid) [below=of high] {Mid-level Model\\(MITRE ATT\&CK™)};
\node (low) [below=of mid] {Low Level Concepts\\(\small{Exploit \& Vulnerability databases})};

\node[absLab] (a) [left=of high] {High};
\node[absLab] (b) [left=of low] {Low};

\draw[<->] (a) -- (b);

\node[absLab1,rotate=90] (arr) [left=20pt of a.north] {\footnotesize{Level of Abstraction}};

\end{tikzpicture}

% \end{document}  
	\caption{Abstraction levels of models and threat knowledge databases \cite{strom_mitre_2018}}
	\label{fig:model-comparison}
\end{figure}
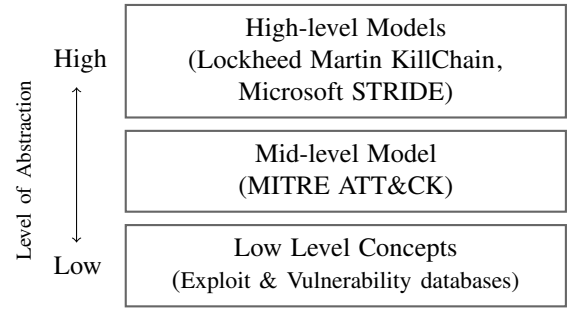

Unlike other models such as Microsoft's STRIDE \cite{shostack_threat_2014} and Lockheed Martin's Cyber Kill Chain \cite{hutchins_intelligence-driven_2011}, \acs{attk} is not highly abstracted from the low level concepts, but at the same time \acs{attk} does not include low level details such as \acf{iocs}, exploits, or vulnerabilities. Fig.~\ref{fig:model-comparison} shows the level of abstraction between high, mid, and low level models. The knowledge base is grounded in observed and plausible adversary behaviours, that are likely to be encountered rather than theoretical techniques that are unlikely to be seen due to difficulty of use or low utility. The behaviours described by the \acs{attk} model can be encoded into IDS systems as signatures, and are also accompanied by potential countermeasures.  

\section{Phineas Fisher}
\label{sec:phineas}

% 4 Aug 2014 - Gamma Group 

Phineas Fisher is a pseudonym \cite{franceschi-bicchierai_hacker_2016}, that identifies as female \cite{fisher_hackback_2019}, who has claimed and verified responsibility for many high profile intrusions and data leaks. In 2014, Fisher targeted Gamma Group \cite{jeff_larson_leaked_2014}. Gamma Group sells surveillance software to governments and police forces around the world, many of which have been criticised by human rights organisations \cite{marquis-boire_bahrain_2012}.  After releasing Gamma Group's client list, source code, and private details, Fisher published a step by step guide \cite{fisher_hack_2014} on how she compromised their systems.

% 5 July 2015 - Hacked Team 

One year later in 2015, Fisher compromised, then published the details and source code for another surveillance company called Hacking Team \cite{j.m._porup_how_2016,franceschi-bicchierai_vigilante_2016}, accompanied with another write up of her methods \cite{fisher_hackback_2016}.

% May 2016 - Catalonia 

In May 2016, she hacked the Catalan police union website \cite{collective_hackback!_2018}, defacing it, then leaked personal information of around 5,000 police officers. Fisher created a video recording of the steps taken in the hack, which showed simple vulnerabilities in their systems. In response to this hack, the police force carried out raids on social centres and hacker labs\footnote{A place for technology hobbyist and enthusiasts to meet. Not related to illegal activities.}, where they claimed they had arrested Fisher. Shortly thereafter Fisher communicated with the media, and agreed to give an interview to Vice News \cite{franceschi-bicchierai_hacker_2016}.

% 19 July 2016 - AKP - Turkey Leak

On the 19th of July 2016, Fisher compromised the Turkish Justice and Development Party (AKP) network \cite{greenberg_wikileaks_2016}, and was collecting data to handover to Wikileaks. While Fisher specifically \cite{emma_best_renowned_2019} told them not to release the data, this was ignored. This hack was not accompanied by a walkthrough guide, and subsequently, Fisher became inactive for a time \cite{franceschi-bicchierai_hacking_2017}. 

% Nov 17 2019 - Social Bug Bounty and Bank hack (2016) confirmation.

In November 2019, Fisher leaked the internal emails of the Cayman Bank and Trust Company located on the Isle of Man \cite{franceschi-bicchierai_phineas_2019}. Along with this leak, she also stole a large sum of money from the bank. This theft has been confirmed, and took place in 2016 \cite{cox_offshore_2019}. As with the other attacks she published a post-mortem \cite{fisher_hackback_2019}, and also offered a 100,000USD bounty to hack banks and oil companies that could lead to the disclosure of documents in the public interest. 
% Who is she ?
To this day no one appears to have been able to identify who Phineas Fisher is, and the Italian investigation into the Hacking Team hack was closed without answers \cite{franceschi-bicchierai_hacking_2018}. While there is some speculation that Phineas Fisher might be a government operation, it is widely believed that she is a hacktivist \cite{franceschi-bicchierai_vigilante_2019}. Fisher's primary message is to start a revolution of hackers, who will hack for the social good, and target companies that are deemed `evil and corrupt'. By publishing her post-mortem documents, she has shown the simple techniques needed to break into these systems. In the case of critical infrastructure such as \acf{ics}, it is therefore valuable to identify how much of a threat these systems might be from hacktivists like Phineas Fisher.

\section{Analysis}
\label{sec:analysis}

This section presents an analysis of Phineas Fisher's intrusions, followed by a discussion of the consequent potential threat to critical infrastructure, and possible detection methods.  

\subsection{Analysis of Fisher's Intrusions}
\label{sub:anlysis-phisher}

The \acl{attk} framework currently has 266 techniques in the enterprise matrix, from these techniques we have chosen a subset which represents Phineas Fisher's tactics and techniques. This is based on her self-published break downs of each attack \cite{collective_hackback!_2018,fisher_hack_2014,fisher_hackback_2016,fisher_hackback_2019}, and is presented in Table~\ref{fig:model}. The table follows the standard \acl{attk} presentation format, where the column headers describe the adversary tactics, while the remaining cells describe the techniques that were performed by Fisher. Each of the tactics are now discussed in turn. Each technique is mapped back to the source: A, Gamma Group; B, Hacking Team; C, Police Union; D, Cayman Bank. Techniques that were not explicitly stated are noted with an `I', which denotes, `Inferred' based on the context. Cells with a red background are mitigated by the countermeasures discussed in section~\ref{subsec:mitigations}. 

\begin{table*}
\caption{A combined \acs{attk} model based on each of Fisher's manifestos.}
\label{fig:model}
\centering
\setlength{\tabcolsep}{2pt}
\begin{adjustbox}{width=\textwidth,angle=90}
\begin{tabular}{@{}L{2cm} L{2cm} L{2cm} L{2cm} L{2cm} L{2cm} L{2cm} L{2cm} L{2cm} L{2cm} L{2cm} L{2cm} L{}}
\toprule 
Initial Access & Execution & Persistence & Privilege \-Escalation & Defense Evasion & Credential \-Access & Discovery & Lateral \-Movement & Collection & Command And Control & Exfiltration & Impact \\ 
\midrule
\fc{a,b,c,d} Exploit Public-Facing Application \m{2} & \fc{i} Command-Line Interface \m{1} & \fc{i} Component
Firmware & \fc{d} Access Token Manipulation \m{5} & \fc{d} Access Token Manipulation \m{5} & \fc{i} Brute
Force \m{4} & \fc{i} Account \-Discovery & \fc{b} Exploitation of Remote Services \m{2} & \fc{a,b,d} Audio
Capture & \fc{c} Commonly Used Port \m{3} & \fc{c} Data Compressed \m{3} & \fc{a} Account Access
Removal\tabularnewline
\f{a,b,d} External Remote Services \m{4}  & \f{i} Execution through API \m{1} & \f{a,b,d} External Remote
Services \m{4} & Exploitation for Privilege \-Escalation \m{2} & \f{i} Clear Command History
& \f{b,c} Credential Dumping \m{5} & \f{i} Application \-Window Discovery & \f{a,b} Logon Scripts \m{7} &
\f{a,b,d} Data from Information Repositories & \f{a,b} Connection Proxy \m{3} & \f{i} Exfiltration
Over Command and Control Channel \m{3} & \f{a,c} Defacement\tabularnewline
\fc{i} Spearphishing Link & \fc{i} Graphical User Interface & \fc{i} File System Permissions
Weakness & \fc{i} File System Permissions Weakness & \fc{i} Component Firmware &
\fc{b,c} Credentials in Files \m{7} & \fc{i} File and \-Directory Discovery & \fc{i} Remote \-Desktop 
Protocol \m{4,5} & \fc{a,b,c,d} Data from \-Local System & \fc{i} Fallback \-Channels \m{3} & \fc{a,b,c} Exfiltration
Over Other Network Medium \m{3} &\tabularnewline
\f{c} Valid Accounts \m{4,5,6} & \f{a,b,d} PowerShell \m{5} & \f{c} Hidden Files and Directories & \f{d} Setuid and
Setgid & \f{a,b} Connection Proxy \m{3} & \f{i} Exploitation for Credential Access \m{2} & \f{a} Network
Service Scanning \m{3} & \f{a,b,c} Remote File Copy \m{3} & \f{a,b} Data from Network Shared Drive &
\f{a,b} Multi-hop Proxy \m{6} & &\tabularnewline
& \fc{i} Scripting & \fc{a,b} Logon Scripts \m{7} & \fc{c} Valid Accounts \m{4,5,6} & \fc{i} Exploitation for Defense
Evasion \m{2} & \fc{i} Forced \-Authentication \m{6} & \fc{a} Network Share Discovery & \fc{i} Remote
Services & \fc{b} Data from Removable Media & \fc{b} Multi-Stage Channels \m{3} &
&\tabularnewline
& \f{i} Source & \f{i} Modify Existing Service & \f{a,b,c} Web Shell & \f{c} Hidden Files and
Directories & \f{a,b,c,d} Input Capture & \f{b} Network \-Sniffing \m{4} & \f{i} Windows Admin Shares &
\f{a,b} Data Staged & \f{i} Multilayer Encryption \m{3} & &\tabularnewline
& \fc{i} Trusted \-Developer Utilities \m{1} & \fc{i} Redundant \-Access \m{3} & & \fc{i} Indirect Command
Execution & \fc{i} LLMNR/NBT-NS Poisoning and Relay \m{6} & \fc{i} Process \-Discovery &
\fc{i} Windows Remote Management \m{5} & \fc{a,b,d} Email Collection \m{4} & \fc{i} Remote Access Tools \m{1,6} &
&\tabularnewline
& \f{i} User Execution \m{3} & \f{i} Registry Run Keys / Startup Folder & & \f{i} Redundant
\-Access \m{3} & \f{b} Network \-Sniffing \m{4} & \f{i} Remote System Discovery & & \f{a,b,c,d} Input Capture &
\f{a,b,c} Remote File Copy \m{3} & &\tabularnewline
& \fc{b} Windows \-Management Instrumentation \m{5} & \fc{i} Security Support Provider & &
\fc{i} Scripting & \fc{b} Two-Factor \-Authentication \-Interception & \fc{i} System \-Information
Discovery & & \fc{a,b,d} Screen Capture & \fc{i} Standard Application Layer Protocol \m{3} &
&\tabularnewline
& \f{b} Windows Remote Management \m{5} & \f{i} Server Software Component & & \f{c} Timestomp &
& \f{a} System Network Configuration Discovery & & & \f{a,b} Standard Cryptographic
Protocol & &\tabularnewline
& & \fc{d} Setuid and Setgid & & \fc{i} Trusted \-Developer Utilities \m{1} & & \fc{i} System Network
Connections Discovery & & & \fc{i} Standard Non-Application Layer Protocol &
&\tabularnewline
& & \f{b} System Firmware & & \f{c} Valid Accounts \m{4,5,6} & & \f{i} System Service Discovery & &
& & &\tabularnewline
& & \fc{c} Valid Accounts \m{4,5,6} & & & & & & & & &\tabularnewline
& & \f{a,b,c} Web Shell \m{5} & & & & & & & & &\tabularnewline
& & \fc{i} Windows \-Management Instrumentation Event Subscription \m{5} & & & & & & &
& &\tabularnewline
\end{tabular}
\end{adjustbox}
\end{table*}

\subsubsection{Initial Access}

In all of the intrusions, initial access was gained by exploiting internet facing applications, typically by performing SQL injection attacks. For the Hacking Team incident, Fisher was able to perform reverse engineering and identify a zero-day vulnerability in their VPN appliance. It later turned out that the appliance was also vulnerable to the trivially performed shellshock\footnote{Shellshock could enable an attacker to cause Bash to execute arbitrary commands and gain unauthorised access to many Internet-facing services, such as web servers, that use Bash to process requests.} vulnerability. While Fisher did not use spear-phishing to gain access, she did refer to them in her guides.

\subsubsection{Execution}

During the time of these exploits, circa 2015, PowerShell was commonly used to perform a lot of execution once initial access had been gained. Today, Microsoft has deployed several mitigations against its misuse, and while this has prevented the same methods from working, there are a plethora of other methods to achieve the same results. This leads on to the other techniques such as Command-Line Interface, Scripting, Graphical User Interface, and  Windows Management Instrumentation/Windows Remote Management, which, if enabled on this target will allow the adversary to execute commands. All of these methods we used or discussed by Fisher. Interestingly, these are all tools normally found in an enterprise network, and follows the philosophy of living off the land, which is strongly advocated in the manifestos. 

\subsubsection{Persistence}

Persistence was often performed using web shells that were uploaded to a compromised service. Hacking Team was a particular exception, where she developed a backdoored firmware for their VPN service. This firmware included many additional tools needed for the next stages. Fisher also maintained a redundant access service, in case she was locked out from her primary persistence method. The guides stated that "I always use Duqu 2 style `persistence', executing in RAM on a couple high-uptime server" \cite{fisher_hackback_2016}, Duqu2 is a relative of Stuxnet, and performed covert, in-memory, espionage operations \cite{maynard_modelling_2016}. 

\subsubsection{Privilege Escalation \& Credential Access}

Privilege escalation was performed by monitoring the activities of operators, using techniques to capture user input and hijack authenticated multi factored sessions, as well as intercepting credentials by modifying popular services to record the plaintext, which was the technique against the Catalonia police union. These approaches are similar to those of state actors.

\subsubsection{Defense Evasion}

In most cases there were few active defences to be evaded, since Fisher tried to maintain a RAM only presence, e.g. exploiting services without placing malware on the disk, which may trigger alerts. When touching the disk, Timestomping was performed, which masks the modification dates of files changed. When impersonating a user login, Fisher would change the logged IP and UserAgent to match historical access logs. 

\subsubsection{Discovery \& Lateral Movement}

All of Fisher's guides start by discovering as much information about the target as possible, typically involving domain and IP scanning for services and other publicly identifiable data. This helps outline the target, and is performed again once an initial compromise has been done. The second time focuses on passive monitoring of network traffic, to find additional targets. Techniques such as LLMNR/NBT-NS poisoning and relay are used which allow for lateral movement. These techniques take advantage of broadcast messages on the network and forge a response to the service to gain an insight into what is running on the network.  

\subsubsection{Collection \& Impact}
Collection and Impact were Fisher's main ATT\&CK tactics (objectives), which was achieved via several techniques. Network file shares were remotely accessed and downloaded locally, with the most common aims being the collection of the target's email archive, internal documentation, client/staff details, and source code. For any company, this can result in a significant impact on the day-to-day operations, and how they are perceived by the public. As a final step, Fisher has previously taken over the company social media account and announced to the world they have been compromised. Although the \acs{attk} model does not have a technique for disclosing private information as an impact tactic, it does include Defacement and Account Access Removal.

\subsubsection{Command And Control \& Exfiltation}
Command and Control, and Exfiltration were performed via commonly used port numbers and connection proxies. While Fisher would use multiple hops and off-the-self remote access tools, and often simple file transfers via HTTP and SSH. These approaches are often sufficient to bypass simple IDS which are unmonitored, as the traffic generated matches day-to-day operations (though more bandwidth may be used, this is often not monitored). 

It is noticeable that Fisher's intrusion methods did not significantly vary between each attack in terms of the techniques used. While the techniques are dependant on the environment, the skills required to perform a successful intrusion are readily attainable.

\subsection{Threat to Critical Infrastructure}
\label{subsec:threat}

Based on the techniques employed by Fisher, we can deduce that a dedicated hacktivist is a valid threat to critical infrastructure. Moreover, in recent years there has been a growing concern for climate change, which may drive people towards targeting oil, gas, and other energy related infrastructure in particular. Such `hacktivist' threats targeting critical infrastructure could feasibly adopt techniques similar to those discussed above, however, the environment found within critical infrastructure is not the same as a traditional enterprise network, due to different underlying operations and requirements. It is common to find older operating systems and applications, which have been validated and certified for specific operations. It may not be possible to update the systems to include the most recent attack mitigations, since that may require additional verification. 
 % powershell mitigations might not be deployed
 % multiple redundant paths may allow an attacker to maintain access
For example, many techniques make use of PowerShell, which was first released in 2006. Since then there has been a great deal of improvements for threat mitigation and event logging. These improvements may not be found within critical infrastructure systems. Moreover, there may be many old Unix systems, and architectures, that contain exploitable vulnerabilities allowing an adversary alternative avenues of attacks. As discovered in our analysis, Fisher would maintain a few remote access paths into their compromised network, to ensure that if one of the compromised machines were detected, she would have another entry point. Within critical networks, there are often multiple redundant network paths providing a resilient network, and while this is a necessity, it also provides adversaries with alternative paths of attack. 

 % spear-phishing 
 % backdoored firmware often seen Duqu/Stuxnet
As reported by the ICS-CERT \cite{ics-cert_incident_2012}, spear-phishing has become common within operators of critical infrastructure. While Fisher did not use this technique, it was mentioned frequently in her manifestos. From our analysis, developing backdoored firmware is within the capability of a hacktivist. This is a concern for critical infrastructure networks as they often contain many embedded devices and network appliances, which may not be recently patched, as was seen in the Duqu and Stuxnet intrusions.  

 % Ukraine - replicated what authorised users would do.
 % data collection, confidential process and plant corp. espionage.
Due to the advancement and proliferation of security controls and mitigations, adversaries are having to resort to more subtle modes of operation. As seen in the 2016 Ukrainian power outage, and by Fisher, the adversary mimicked legitimate users actions to avoid detection. The motivation of a hacktivist might be to find and leak information about the company or to disrupt operations. Leaking information could be a concern for manufacturing companies, which often have trade secrets encoded into the network. Meanwhile, power generation and transmission operators may have financial fines imposed for service disruptions. 

\subsection{Mitigations}
\label{subsec:mitigations}

Based on the analysis of the tactics and techniques used by Fisher, which could potentially be deployed by anyone hacktivist threat actor, we now highlight seven mitigations methods defined by the \acs{attk} framework, that may be deployed within critical infrastructure systems. The mitigations are ordered by level of deployment complexity, and were chosen based on the number of techniques which they mitigate: 
 
\begin{enumerate}
  \item \underline{Execution Prevention}: Application whitelisting may be able to prevent the running of executables masquerading as other files.
  \item \underline{Application Isolation and Sandboxing}: Perform application isolation via operating system calls, or virtualisation and application microsegmentation to mitigate the impact of a compromise.
  \item \underline{Network Intrusion Prevention}: Network intrusion detection and prevention systems that use network signatures to identify traffic for specific adversary malware can be used to mitigate activity at the network level. Signatures are often for unique indicators within protocols and will be different across various malware families and versions. Adversaries will likely change tool signatures over time or construct protocols in such a way as to avoid detection by common defensive tools. In which case anomaly based IDS may be used.
  \item \underline{Multi-factor Authentication}: Integrating multi-factor authentication (MFA) as part of the organisational policy can greatly reduce the risk of an adversary gaining control of valid credentials that may be used for additional tactics such as initial access, lateral movement, and collecting information. MFA can also be used to restrict access to cloud resources and APIs.
  \item \underline{Privileged Account Management}: Audit account and group permissions to ensure that accounts used to manage servers do not overlap with accounts and permissions of users in the internal network that could be acquired through Credential Access and used to log into the Web server and plant a Web shell or pivot from the Web server into the internal network.
  \item \underline{Filter Network Traffic}: Use host-based security software to block non essential traffic e.g. LLMNR/NetBIOS.
  \item \underline{Restrict File and Directory Permissions}: Restrict write access to scripts to specific administrators. Where possible perform access and execution logging.
\end{enumerate}

Table~\ref{fig:model} includes which of the mitigations may prevent each techniques, by colouring the cells red and including a number of each mitigation. 

\section{Conclusion}
\label{sec:conclusion}

As far as the authors are aware this is the first academic analysis of Phineas Fisher, and the first paper to provide a technical analysis of the `hacktivist' threat to critical infrastructure. We have taken a previously unknown threat actor and identified a set of tactics and techniques which may be used to mitigate future attacks. We are in the process of submitting this threat actor into the \acf{attk} knowledgebase, which will be available to other researchers and security practitioners.  More broadly, research is also needed to detect and prevent such threat actors within the industrial control landscape.

\section*{A note on reproducibility}

All information used in the creation of these models are cited in the main body of the text. Since some of the manifestos were difficult to ascertain we maintain a local copy\footnote{\url{https://github.com/PMaynard/Viva-Phineas-Fisher}}, which includes the individual \acs{attk} models as well as the combined model discussed in this manuscript.

\section*{Acknowledgements}

The authors wish to thank the reviewers for their helpful feedback. We also wish to extend our thanks to the hosts of the Risky Biz podcast (Patrick Gray and Adam Boileau), who provided enlightening reports into Fisher's exploits and brought Fisher to the authors' attention. 

\bibliographystyle{IEEEtran}
{\small
\bibliography{IEEEabrv,phisher}}

\end{document}